\documentclass{svjour2_}                    % onecolumn
\smartqed  % flush right qed marks, e.g. at end of proof
\usepackage{graphicx}
\usepackage{color}
\usepackage[colorlinks,bookmarksopen]{hyperref}
\usepackage{amsmath}
\begin{document}

\title{Opinion dynamics with disagreement and modulated information %\thanks{Grants or other notes
}


\author{Alina~S\^irbu \and 
Vittorio~Loreto \and 
Vito~D~P~Servedio  \and 
Francesca Tria
}


\institute{A. S\^irbu \at
              ISI Foundation, Via Alassio 11c, 10126 Turin, Italy \\
              \email{alina.sirbu@isi.it}           
           \and
           V.Loreto \at
				Sapienza University of Rome, Physics Dept., P.le~A.~Moro 2,
        00185 Rome, Italy\\
ISI Foundation, Via Alassio 11c, 10126 Turin, Italy \\
              \email{vittorio.loreto@roma1.infn.it}
\and
			V.D.P. Servedio \at
				Sapienza University of Rome, Physics Dept., P.le~A.~Moro 2, 00185 Rome, Italy\\
              \email{vito.servedio@roma1.infn.it}
			  \and
			  F. Tria \at
			  ISI Foundation, Via Alassio 11c, 10126 Turin, Italy \\
              \email{tria@isi.it} 		  
}

\date{}

\maketitle

\begin{abstract}
Opinion dynamics concerns social processes through which populations
or groups of individuals agree or disagree on specific issues. As
such, modelling opinion dynamics represents an important research area
that has been progressively acquiring relevance in many different
domains. Existing approaches have mostly represented opinions through
\emph{discrete} binary or \emph{continuous} variables by exploring a
whole panoply of cases: e.g.\ independence, noise, external effects,
multiple issues. In most of these cases the crucial ingredient is an
attractive dynamics through which similar or similar enough agents get
closer. Only rarely the possibility of explicit disagreement has been
taken into account (i.e., the possibility for a repulsive interaction
among individuals' opinions), and mostly for discrete or 1-dimensional
opinions, through the introduction of additional model
parameters. Here we introduce a new model of opinion formation, which
focuses on the interplay between the possibility of explicit
disagreement, modulated in a self-consistent way by the existing
opinions' overlaps between the interacting individuals, and the effect
of external information on the system. Opinions are modelled as a
\emph{vector of continuous} variables related to \emph{multiple
  possible choices} for an issue. Information can be modulated to
account for promoting multiple possible choices. Numerical results
show that extreme information results in segregation and has a limited
effect on the population, while milder messages have better success
and a cohesion effect. Additionally, the initial condition plays an
important role, with the population forming one or multiple clusters
based on the initial average similarity between individuals, with a
transition point depending on the number of opinion choices.

\keywords{Opinion dynamics \and interaction \and disagreement \and external information \and numerical simulations }
\end{abstract}

\section{\label{sec:intro}Introduction}
Opinion formation is an important property of social
systems. People are driven by their choices in many moments of their
lives, based on opinions formed in time under the influence of many
factors, such as their own personality, the culture they belong to,
peer interaction, mass-media effects, etc. These choices range from
selecting a lifestyle, a specific behaviour, a phone company, or a
supermarket, to the city to leave in or whom to vote for. The way
opinions form and choices are made is of interest to many classes of
researchers, with implications in many fields from politics to
economics, to marketing. Probably the most immediate example is
represented by political opinions: here the question is how people
synthesize the different sources of information and biases, to come up
with a specific position to be eventually expressed as a vote whenever
required. Another important example concerns marketing strategies.
These strategies are typically devised using social research, under
the assumption that in order to attract clients, the products have to
be presented in an appealing way. An example is described in
\cite{duhigg2012power}, where a supermarket chain, after careful
analysis of the purchases of their clients, send customised
promotional leaflets. By sending early advertising about products
people need, they attract them before other chains, and increase sales
also in departments other than that advertised. For a different client
type, the advertising looks different, even though the supermarket is
the same. This is one classical example of shaping external
information and advertising in such a way as to attract as many
individuals as possible. Yet another example is the decision to buy a
specific product in a market segment, e.g.\ a mobile phone, a car, a
book, etc. It is again a matter of opinion formation and evolution,
based on an evaluation of how the product matches the needs, attitude
and possibilities of an individual. For instance, while until a few
years ago, the smart phone market was very narrow, nowadays it has
extended significantly \cite{techRev}. It would be interesting to know
what changes people's opinion, and leads to a large propagation of a
new technology.

Traditionally studied by sociologists, opinion formation has become an important topic for physicists and different classes of models have been investigated in the past years \cite{Castellano2009a}. Models with opinions or options modelled through discrete variables, such as the voter, local majority rule \cite{Galam2008a}, Sznajd \cite{Sznajd-Weron2001}, Axelrod \cite{Axelrod1997}, social impact \cite{Lewenstein1992,Nowak1996,vallacher2008} and their various extensions, have been applied to explain aspects of elections, strikes, dynamics of mobile markets, changes in the number of privately owned companies, financial crises and culture formation \cite{Galam2010,Galam2011,Lima2012,nowak2000}. Although discrete variables are very suitable for modelling choices, the internal state of individuals, based on which discrete decisions are taken, may be continuous. Also, the opinion itself could take continuous values. The Deffuant-Weisbuch \cite{Deffuant2000,Weisbuch2002} and Hegselmann-Krause \cite{Hegselmann2002} models analyse one-dimensional continuous opinions, while multiple dimensions are analysed in \cite{Lorenz2007}, where the opinion space is the unit simplex. The Continuous Opinions and Discrete Actions approach (CODA) \cite{Martins2008a} analyses internal probabilities for two or three discrete choices. Most of the original models deal with attractive dynamics, where individuals follow their neighbours. However, in reality, both attractive and repulsive interactions can be observed, and it has been argued that disagreement is a very important feature of a democratic society \cite{Huckfeldt2004}. This element has been introduced in some of the existing models, by considering additional model parameters to control disagreement (e.g.\ \cite{Radillo-Diaz2009,VazMartins2010,Kondrat2010,Sznajd-Weron2011,Nyczka2012,Kurmyshev2011,Hong2011,Acemoglu2010}). Additionally, the effect of external information is crucial when modelling real social systems. Again, some approaches consider this effect \cite{Carletti2006,Gargiulo2008a,Peres2011,VazMartins2010,Crokidakis2011a,Hegselmann2006}. However, an analysis for multi-dimensional continuous opinions is missing, both for external information and disagreement.

Here, the interplay between disagreement and external information in opinion dynamics is analysed, by the introduction of a new modelling approach. The model considers the probabilities that an individual will make a specific choice out of multiple possibilities (such as voting or choosing a market product). It includes both attractive and repulsive interactions, in a self consistent way, without the addition of a further parameter to the model. Modulated information is also possible, which promotes multiple choices at the same time.  Additionally, an analysis of the effect of the initial condition is performed.

\section{Methods} 
\begin{figure*}
\centering
\includegraphics[width=0.7\textwidth]{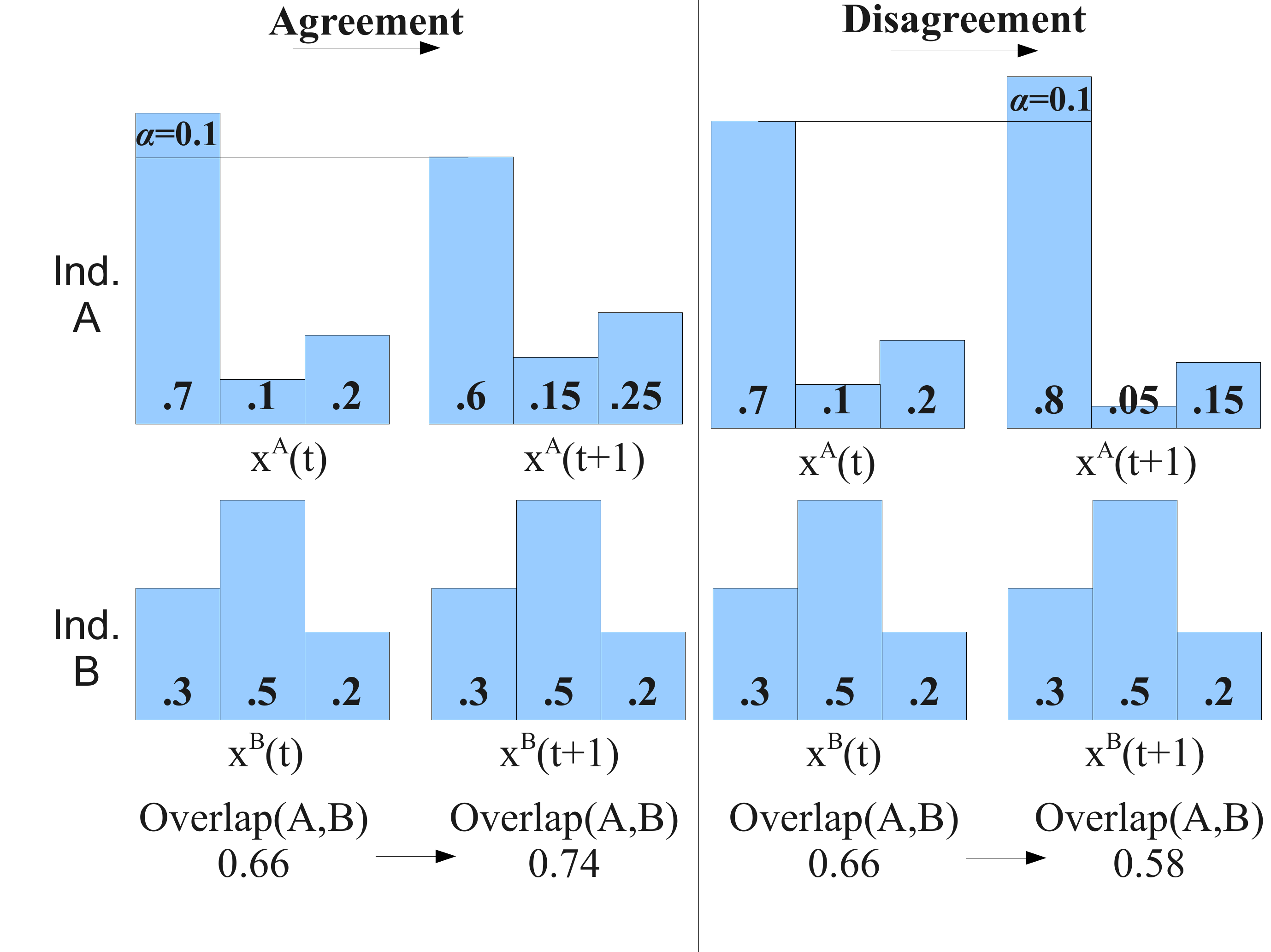}
\caption{\textbf{Example of pairwise interaction.} 
%  \vito{nell'istogramma in alto a destra cambiare il .6 in .8}
  We consider a generic
  interaction between an individual A and an individual B, in a
  dynamics where the number of possible opinions is fixed to $K=3$. In
  particular, the individual A changes its opinion according to its
  interaction with the individual B.  We depict both the cases of
  agreement (left) and disagreement (right). 
  Left: The individual A
  changes its status according to an agreement with individual B. In
  this case, the opinion they discuss is the opinion $l=1$, and its
  probability is decreased by the value $\alpha=0.1$ in the individual
  A to become more similar to the individual B (and, as such, the
  mutual overlap increases).  The probabilities of the rest of the
  opinions are increased by equal amounts (in this case, $\frac{\alpha}{2}$) to
  conserve the unity sum. Right: the same dynamics in case of
  disagreement. In this case, the individual A increases the
  probability of opinion 1 by the amount $\alpha=0.1$, decreasing its
  overlap with the individual B. Equivalently, the probabilities of
  the rest of the opinions are decreased by equal amounts to conserve the unity sum. 
%  \vito{che succederebbe se fosse $p^A_k < \alpha/2$? Perch\'e non abbiamo usato $p_k/\sum p_k$ come normalizzazione?}
}
\label{fig:update}
\end{figure*}

\noindent A fully connected social network of $N$ individuals is
considered, where each agent has to make a choice between $K$ possible
opinions on a given subject. Each individual maintains a set of
probabilities for the $K$ possibilities: $\vec x=[p_1,p_2,\ldots,p_K]$ with
%
%\begin{equation}
\(
  \sum_{k=1}^{K}{p_k}=1
\)
%\end{equation}
%
i.e., an element in the simplex in $K-1$ dimensions. We define the
similarity between two individuals $i$ and $j$ as the cosine
overlap between the two opinions' vectors: 
\begin{equation}
  o^{ij}=
    \frac{\vec{x}^{\,i}\cdot\vec{x}^{\,j}}{|\vec{x}^{\,i}| |\vec{x}^{\,j}|} =
    \frac{\sum_{k=1}^{K}{p_k^i p_k^j}}{\sqrt{\sum_{k=1}^{K}{(p_k^i)^2} \sum_{k=1}^{K}{(p_k^j)^2}}}.
%  o^{ij}=
%    \frac{\sum_{k=1}^{K}{x_k^i x_k^j}}{\sqrt{\sum_{k=1}^{K}{(x_k^i)^2} \sum_{k=1}^{K}{(x_k^j)^2}}}.
\end{equation}
At each time step a randomly selected pair of individuals, ($i$,$j$), interacts, either by agreeing or disagreeing, based on their instantaneous overlap:
\begin{equation}
  p_\mathrm{agree}^{ij}=\min(1,\max(0,o^{ij}\pm\epsilon)) \label{eq:agree},
  \end{equation}
  \begin{equation}
    p_\mathrm{disagree}^{ij}=1-p_\mathrm{agree}^{ij} \label{eq:disagree}.
  \end{equation}
 where $\epsilon$ is
  a noise term which avoids lack of interaction due
  to null overlap, with the choice between the plus and minus signs (Equation \ref{eq:agree}) made randomly at each time step.
  The rule~\ref{eq:agree} (and~\ref{eq:disagree}) corresponds to imagine that, during the interaction, each
  individual perceives how close (or distant) he/she is from another
  individual and consistently agrees or disagrees. It is important to note here that our model does
  not impose  ``bounded confidence'': individuals with low overlap will tend to
  disagree, with an effect on their state, and it is however possible
  that two individuals agree even though their overlap is null (with low
  probability). Furthermore, the model introduces the possibility of
  both agreeing and disagreeing in a self consistent way. Interaction
  causes one of the individuals in the pair to change a random element
  $l$ in the opinion vector:
  \begin{equation}
	p^i_l(t+1)=
	\begin{cases}
		p_l^i(t) \pm \alpha\, \mathrm{sign}(p_l^j-p_l^i) & \text{if } |p_l^j-p_l^i|>\alpha
		\\ 
		p_l^i(t) \pm \frac{1}{2}(p_l^j-p_l^j) & \text{otherwise.}
	\end{cases}
\label{interEq}
\end{equation}
where the plus sign occurs when the interaction results in agreement and
the minus sign when the interaction results in disagreement.  Hence, the
position is changed by  a small fixed step $\alpha>0$, unless
differences to the other individual are too small, in which case the
change is half the difference. This allows complete agreement between
individuals. The parameter $\alpha$ determines the flexibility of
agents since it fixes the time scale for local agreement or
disagreement. The larger $\alpha$ is, the faster the two individuals will agree or
get separated.  The rest of the elements in $\vec x^{\,i}$ are adjusted to
preserve the unit sum, by uniformly redistributing the amount the
element $l$ was changed by. Since $0$ and $1$ are absorbing values, this is performed iteratively. For instance, if we consider that position $l$ has been increased by $\alpha$, the other positions would have to change by $-\frac{\alpha}{K-1}$. In attempting to do this, some positions may become negative. In this case, the negative positions will be set to $0$, after being summed to obtain a new amount $\alpha'$ for redistribution. This new amount will be redistributed to all non-null positions (except for $l$), with the procedure repeated until no negative position is obtained. This method allows for the absolute value of the change on position $l$ to be the same for agreement and disagreement. Figure \ref{fig:update} demonstrates
graphically the update rule employed.\\
External information, e.g.\ mass-media, is introduced as a static agent
$\vec I=[I_1,I_2,\ldots,I_K]$ with:
%
%\begin{equation}
\(
\sum_{k=1}^{K}{I_k}=1
\).
%\end{equation} 
%
After interacting with a peer, an individual interacts also with the
information with probability $p_I$, following the same interaction
rules. Hence, interacting with the external information does not imply
less peer communication. In previous models, external information
biased individuals towards one choice out of all possibilities (e.g.\ Axelrod, Sznajd). 
Here, this means setting one position in $\vec I$ to 1
and others to 0. In reality, however, sources of information are so
wide that only one possibility is never promoted. Our approach has the
ability to model such complex influence, by choosing non zero values
on more positions of $\vec I$, i.e., a \emph{modulated information}. This
was also true for the Deffuant model: if we consider that the
continuous opinion is actually a probability to make a choice between
two discrete options, then milder external information can be
introduced by using information values far from the extremes of the
opinion interval. In fact, similarities between information effects in
the Deffuant model and the one introduced here will be discussed in a
later section.

\section{Results}
\subsection{Role of the initial condition}\label{initSect}

\begin{figure*} \centering
  \includegraphics[width=\textwidth]{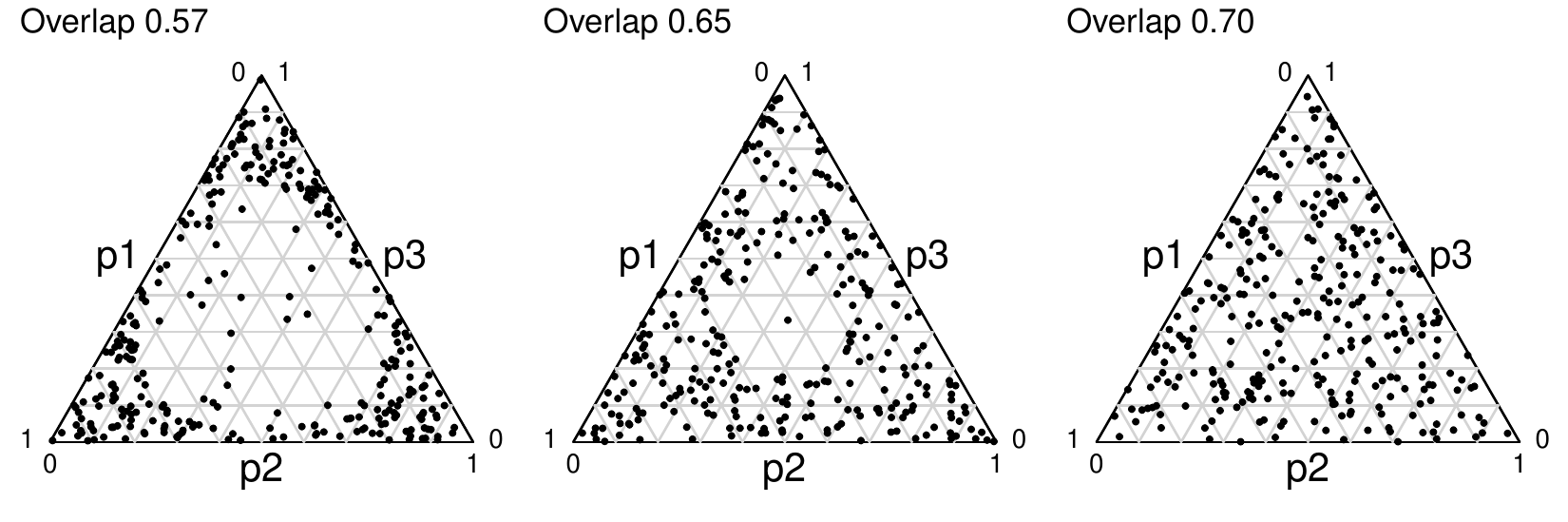} \caption{Initial
    conditions for $K=3$ - distribution of opinion values for a
    population of 300 individuals, with different average population
    overlap. The entropy threshold values used to generate these populations are, from left to right, $1.2$, $1.5$ and $1.6$.  }
\label{init}
\end{figure*}

An important parameter to take into account is the initial average
overlap of the population, defined as
:\begin{equation} 
	\bar{o}=\frac{2\sum_{i,j}o^{ij}}{N(N-1)}.
\end{equation}
This value represents the probability that a randomly chosen pair of
individuals will follow agreement dynamics, so it may have a large
influence on the final state of the population. Thus, it is
interesting to see how the dynamics depends on this feature of the
initial condition, and we perform this analysis when no external
information is present ($p_I=0$). A random sampling of the simplex in $K-1$
dimensions yields a population with a relatively large average
overlap. In order to generate populations with different initial
$\bar{o}$, we consider the entropy $S$ associated to the opinions
probability, defined as: 
\begin{equation} 
S=-\sum_{i=1}^{K}{p_i  \log_2(p_i)} 
\end{equation}
and we remove from the random sampling some individuals that have $S$
larger than a specific threshold. Specifically, we construct a
population of $N$ individual by random sampling each individual from a
simplex in $K-1$ dimensions. In order to decrease the population
average overlap, we compute for each sampled individual its entropy
$S$ and remove it from the population with probability $0.9$ if $S$ is
above the chosen threshold. We continue to sample points in the $K-1$
simplex and to apply the above procedure till we reach a population of
the desired size $N$. Decreasing the threshold, populations with
decreasing $\bar{o}$ can be obtained. Figure \ref{init} shows the
distribution of populations with different $\bar{o}$ for $K=3$.
Throughout the rest of this paper, more fragmented or compact
population will be generated in this manner, with the most compact
initial condition corresponding to the random sampling of the simplex.

In order to study the effect of the initial condition, we have
performed numerical simulations ($N=300$, $\epsilon=0.1$, $p_I=0$) for
different $K \in \{3,5,10,20,$ $30\}$, with corresponding $\alpha
\in\{0.0167,0.01,0.005,0.0025,0.00167\}$. Hierarchical clustering of
the final population has been performed, using complete linkage
clustering \cite{manning1999}, and cutting the tree at 0.8 similarity level. This ensures
that if two agents $i$ and $j$ are in the same cluster then
$o_{ij}>0.8$. The {\em effective} number of clusters has been computed
as the cluster participation ratio (PR):

\begin{equation}
PR=\frac{\left(\sum_{i=1}^{C}{c_i}\right)^2}{\sum_{i=1}^{C}{c_i^2}}
\end{equation}
with $C$ the number of clusters and $c_i$ the size of cluster $i$.
This measure is more significant than the number of clusters itself,
as it also considers cluster sizes. For instance, if the population
consists of two clusters, PR would be 2 only if the clusters are equal
in size, and very close to 1 if one of the clusters is extremely small
compared to the other.

\begin{figure*}
\centering
\includegraphics[width=0.7\textwidth]{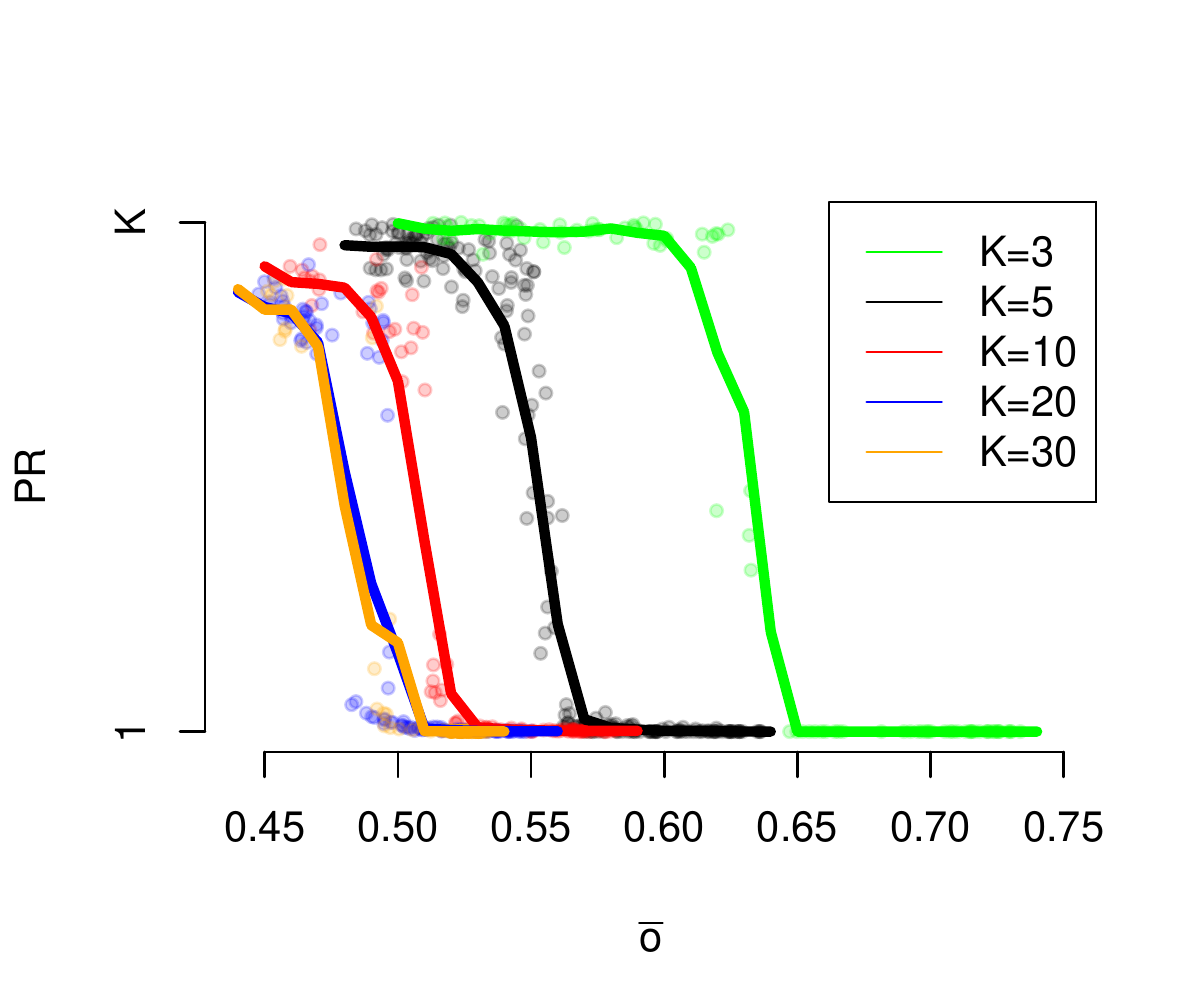}
\caption{Effect of initial condition on the effective number of
  clusters in the final population. Dots represent individual
  instances, while lines are
  averages. Simulations have been performed with $N=300$, $\epsilon=0.1$, $p_I=0$ for
different $K \in \{3,5,10,20,30\}$, with corresponding $\alpha
\in\{0.0167,0.01,0.005,0.0025,0.00167\}$.  }
\label{initTransitions}
\end{figure*}

Figure \ref{initTransitions} displays PR values for different
realizations of the model, depending on the initial $\bar{o}$. This
shows that the initial condition has a large effect on the final
population, with a transition in the number of clusters obtained.
Specifically, above a certain value of the initial overlap, the
population forms one cluster, while below this value the population
divides into K clusters, each of the form (1,0,\ldots). The value of
$\bar{o}$ marking the transition decreases with $K$, showing that
agreement in the population is facilitated by the existence of more
opinion choices. This has also been observed for the model in
\cite{Lorenz2007}, although without performing an analysis of the
initial condition effect. However, it is important to note that in the
case of agreement (and in the absence of external information), agents
maintain a probability different from zero for (almost) all options,
which means a generalised state of indecision.
On the other hand, when clusters form, these adopt a more decided
option with $p_l=0$ for many values of $l$. More details about the structure of the opinions
 in the different cases will be discussed in Section \ref{sec:temporal}.

\subsection{Effect of information}

We now analyse the effect of external information on the final state
of the population. Numerical simulations have been performed for
$N=300$ $K=5$, $\alpha=0.01$, $\epsilon=0.1$ and $p_I$ ranging from 0
to 1. Four types of information have been investigated, in order to
study the effect of extreme and mild external messages on the
population (specifically $\vec I \in
\{[1,0,0,0,0],[0.8,0.2,0,0,0],[0.4,0.2,$ $0.2,0.1,0.1],[0.2,0.2,0.2,0.2,0.2]\}$).
Since, as shown in section \ref{initSect}, the initial condition plays
a very important role in the number of clusters obtained, four
different initial conditions have been used, with
$\bar{o}\in\{0.49,0.55,0.57,0.62\}$. These values cover situations
before, around and after the transition occurs (Figure
\ref{initTransitions}), with $\bar{o}=0.62$ corresponding to the
random sampling of the simplex.

The effective number of clusters has been determined as in the
previous section, using the PR measure. The effect of information has
been quantified by computing the average information overlap
($\overline{IO}$) once the population has reached a stationary state:
\begin{equation}
\overline{IO}=\frac{1}{N}\sum_{i=1}^{N}{o^{Ii}}
\end{equation}
where $o^{Ii}$ represents the cosine overlap between agent $\vec x^{\,i}$ and
the external information $\vec I$. This average measure is an indicator of
the percentage of individuals in the population adhering to the
information.
 
\begin{figure*} \centering
  \includegraphics[width=\textwidth]{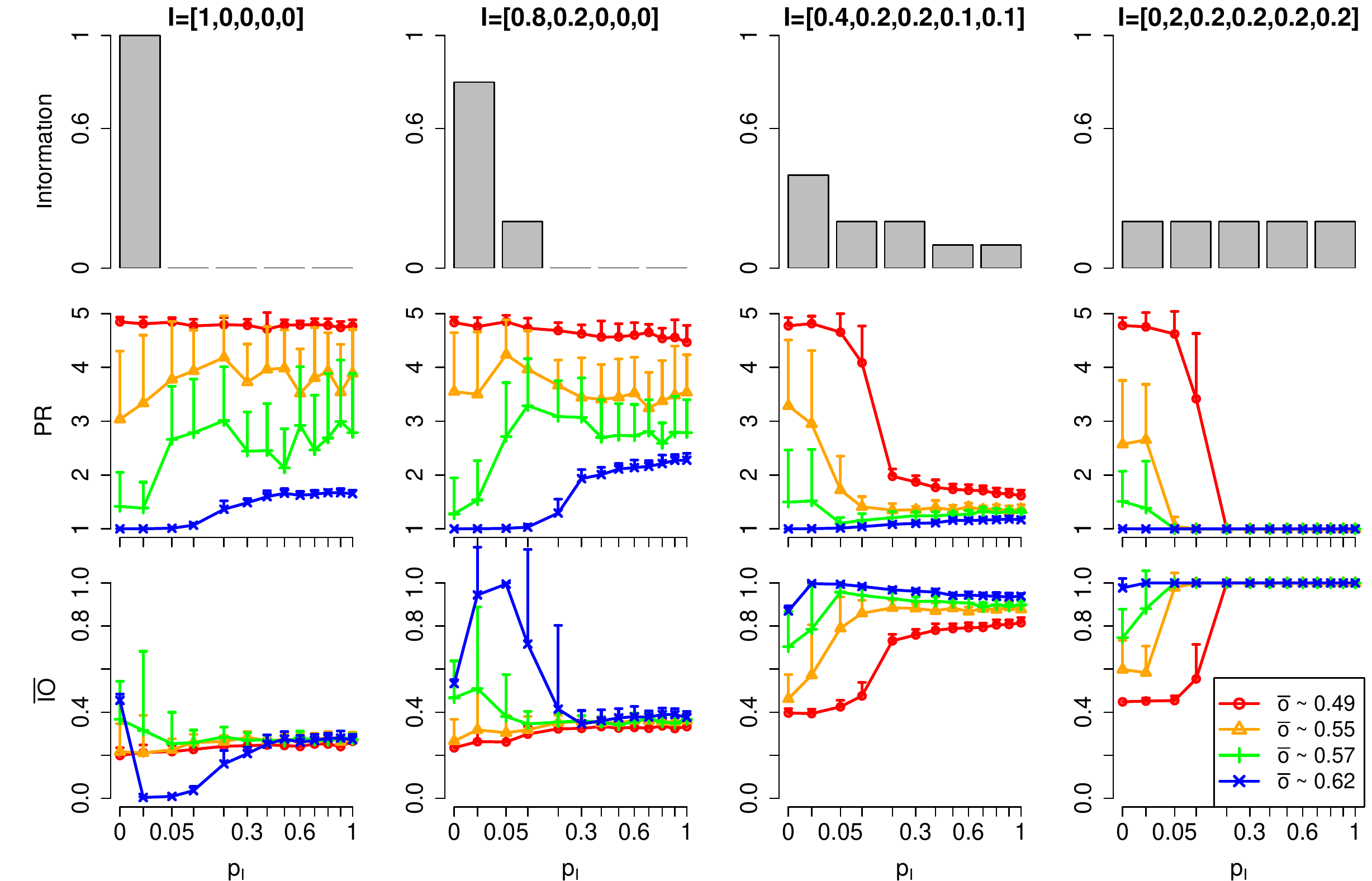}
  \caption{Numerical results for a population of 300 individuals,
    $K=5$, $\epsilon=0.1$ and $\alpha=0.01$ are presented (100000
    population updates (refer to figure~\ref{times} for the time
    needed for a population of size $N$ to reach a stationary state),
    20 instances for each parameter value). Four different information
    values are used, corresponding to each column. The top graphs show
    the histogram of the information, the middle the average number of
    clusters as a function of $p_I$ while the bottom the average
    information overlap again as a function of $p_I$. The individual
    points correspond to
    $p_I\in\{0,0.01,0.05,0.1,0.2,0.3,0.4,0.5,0.6,0.7,0.8,0.9,1\}$.
    Error bars are displayed as well, showing standard deviation for
    each point. Where error bars are not visible, the standard
    deviation is within graphic point limits or null. }
\label{info5}
\end{figure*}
Figure \ref{info5} displays $\overline{IO}$ and the effective cluster
number PR for the different parameter configurations, each point being
an average over 20 realizations of the process. Several patterns can
be observed. In general, extreme information is less successful in the
population compared to milder messages, as $\overline{IO}$ values
indicate. Additionally, mild information favours cohesion in the
population, i.e., a decreased number of clusters compared to the
$p_I=0$ situation, while extreme information induces segregation
(increased PR). These effects increase with $p_I$. Of course, the
extent of the two effects observed depends on the initial condition.
That is, the segregation effect of extreme information is small when
the population starts from a tight community, i.e., large $\bar{o}$,
with the number of clusters increasing when $\bar{o}$ decreases.
Similarly, the success of mild information is smaller when the initial
population is not very compact and larger in the opposite case.

Several other interesting details can be observed. In general, even
when the frequency of interaction with the external information
($p_I$) is very large, the success in the population ($\overline{IO}$)
is bounded. This bound depends both on the initial condition and on
the type of information, with a small value for extreme information
and low initial $\bar{o}$, and (nearly) complete success for mild information
and large $\bar{o}$. This is due to the disagreement dynamics based on
the overlap between individuals and the external information, and is
observable also in real life, where no matter how much propaganda
there is, if the individuals do not agree enough with an idea, this is
not adhered to.

\begin{figure*} \centering
  \includegraphics[width=\textwidth]{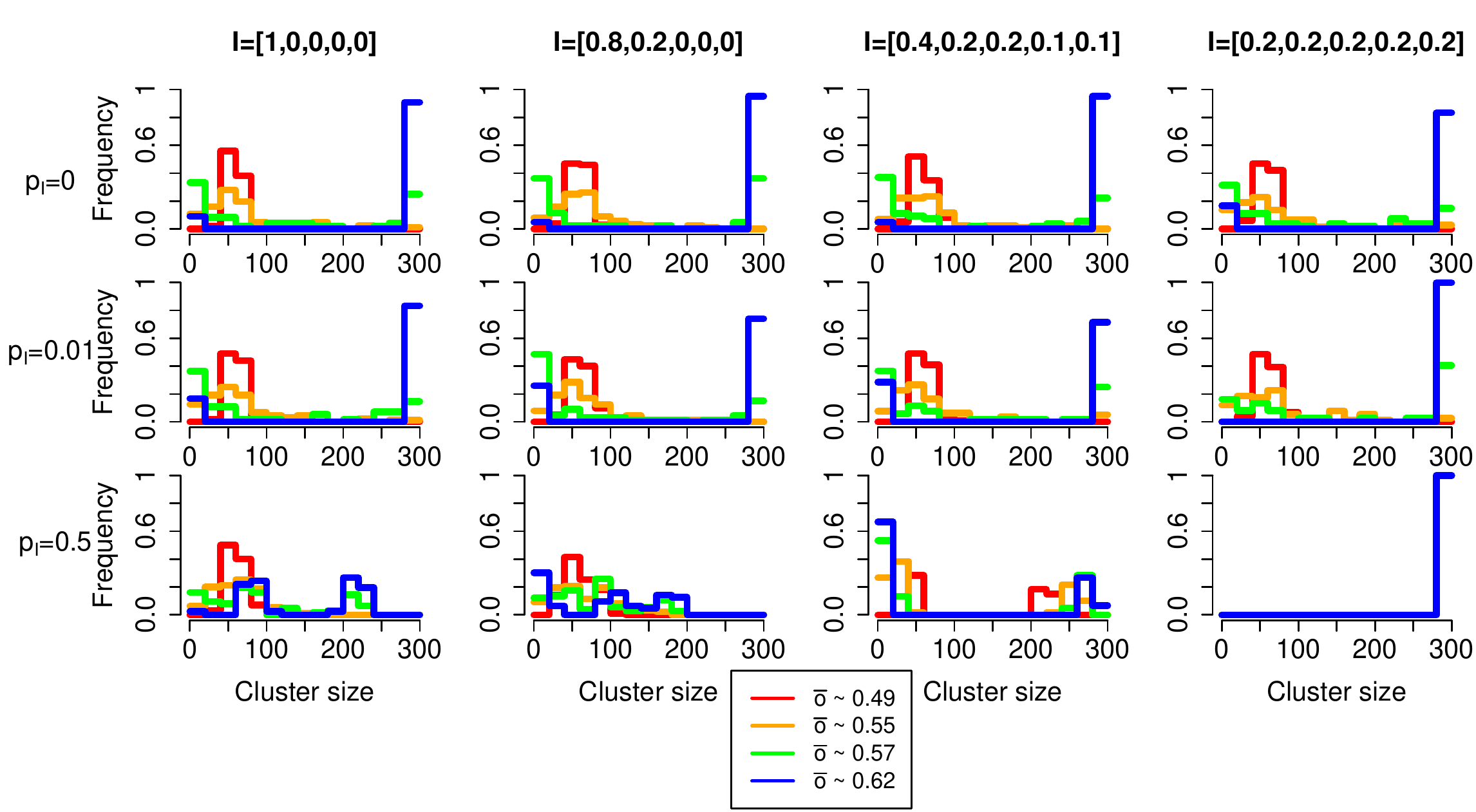}
  \caption{Histograms of cluster sizes obtained over 20 simulations
    runs for $K=5$, $N=300$, $\epsilon=0.1$ and $\alpha=0.01$. Four
    information types and four initial conditions are displayed, for
    $p_I \in \{0,0.01,0.5\}$. }
\label{clusterSizes5}
\end{figure*}

To analyse in more detail the clustering patterns obtained, Figure
\ref{clusterSizes5} displays histograms of cluster sizes obtained in
20 runs, for $p_I\in\{0,0.01,0.5\}$. This shows that, in general, as
expected, PR values of 5 correspond to clusters of similar size
(around 60 individuals), while PR values close to 1 are generated by
one very large cluster and one or several extremely small groups. The
figure shows very clearly how the fraction of very large cluster sizes
increases as the initial condition becomes more compact (red to blue
lines) and as the information becomes milder (left to right columns).

Another very interesting phenomenon can be observed for compact
initial conditions, i.e., random sampling of the simplex (Figures
\ref{info5} and \ref{clusterSizes5}, blue lines). When the information
is not too extreme $\vec I=[0.8,0.2,0,0,0]$, the entire population adheres
to it provided $p_I$ is very small, while as $p_I$ increases, the
media success decreases. On the other hand, when the information is
very peaked ($\vec I=[1,0,0,0,0]$), very small $p_I$ leads to complete
disagreement to the population, while larger $p_I$ increases
agreement. Cluster sizes, however, do not show a large change when
increasing $p_I$ from $0$ to $0.01$, showing that group composition is
basically the same, even though $\overline{IO}$ values change
significantly. This indicates that a very low $p_I$ allows for the
dynamics to proceed in a similar manner as without information, and
after groupings are formed, these are slowly swayed towards or away
from the information. In the specific case here, a compact initial
population forms first one cluster, which moves close to the
information if this is mild, or far from it if peaked. This shows
that, when facing a compact group, an external message is more
efficient if presented gradually, provided it is not extremely
different from the current convictions of the population. This
suggests that peer influence is more effective than that from an
external static source. This can be explained by the fact that peers
are flexible, and move towards others freely, while external
information is too rigid. An agent that is facing an external message
which she does not agree with, will move away from it, while when the
direct exposure to the external message is small, the interaction with
other peers can sway her towards accepting the message. However, for
this to happen, the message has to be close enough to the initial
state of the population, i.e., acceptable by a large number of
individuals. When this is not true, as is the case with an extreme
external information, the entire population will disagree with the
message, and the media campaign will have no effect.

\begin{figure*} \centering
  \includegraphics[width=\textwidth]{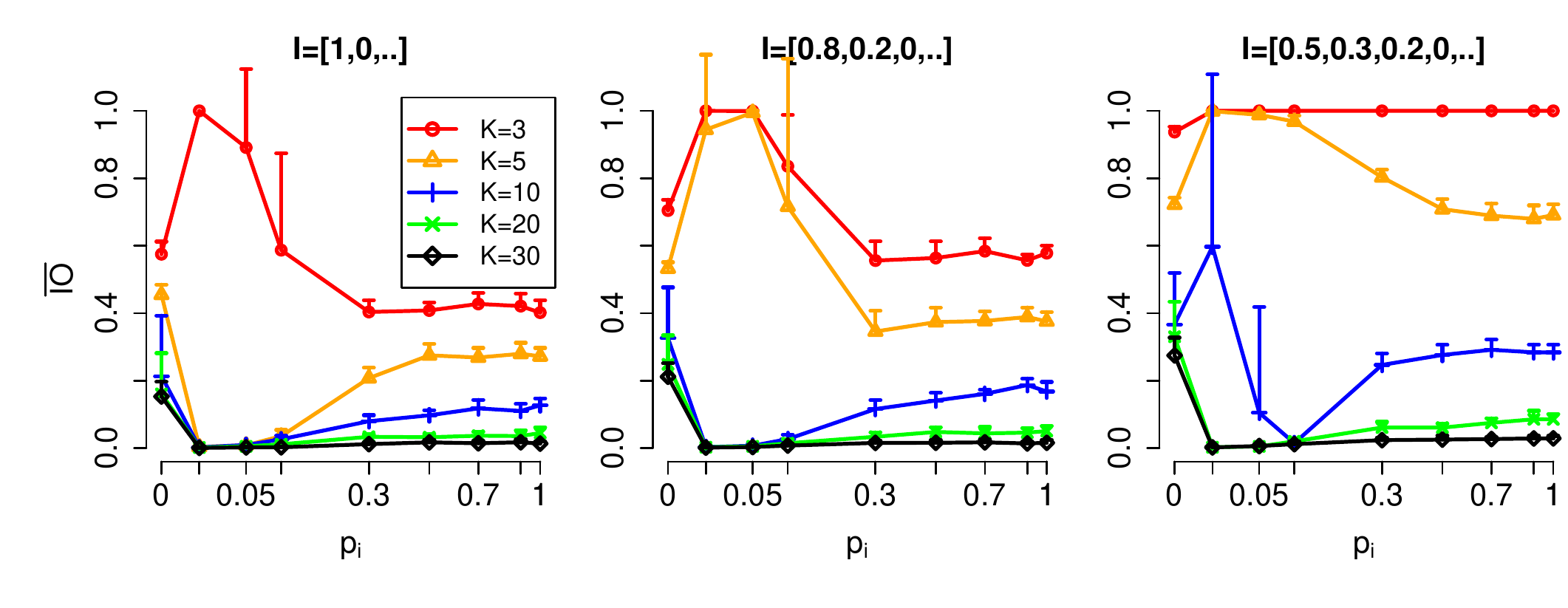}
  \caption{ Average information overlap for compact populations of 300
    individuals, $K\in\{3,5,10,20,30\}$, $\epsilon=0.1$ and corresponding $\alpha
    \in \{0.0167,0.01,0.005,0.0025,0.00167\}$ (values for $\alpha$ have been scaled to maintain the scale, compared to the average value of $p_i$, when $K$ increases). The graph includes 10 instances for each
    parameter value. Error bars show one standard deviation from the
    plotted mean. The individual points correspond to
    $p_I\in\{0,0.01,0.05,0.1,0.3,0.5,0.7,0.9,1\}$.}
\label{infoK}
\end{figure*}

It is important to note that the overlap of the external information
with the population, which, as our results show, is one of the most
important determinant for the success of a campaign, depends also on
$K$. For instance, an extreme message $[1,0,\ldots]$ has an average
overlap with a random population of $0.447$ when $K=5$ and $0.577$ for
$K=3$. Figure \ref{infoK} shows average overlap with the information
obtained for different $K$ values. Each initial population has been
obtained by random sampling of the simplex, being thus a compact
population. Three information types have been used. It is obvious that
the three types of information have a very different effect depending
on $K$. While for $K=3$, information $[0.5,0.3,0.2]$ is very mild,
having complete success, for $K=30$ $[0.5,0.3,0.2, 0,\ldots, 0]$ would
be quite extreme, since only three out of thirty options are promoted,
so the information overlap obtained decreases drastically. We can
conclude that the success of all three information types in the
population decreases with $K$. This indicates that it is easier to
convince the public about a specific option when there are few
choices, compared to when the number of choices is large. When $p_I$
is small ($\neq0$), the phenomenon explained in the previous paragraph for compact populations can be
observed, for all $K$.  Specifically, when information is mild enough, complete agreement in the population is obtained ($\overline{IO}=1$),
while extreme information fails completely in attracting individuals ($\overline{IO}=0$). Again, mildness of information depends on $K$. For example, for low $p_I\neq0$, information $[0.8,0.2,\ldots]$ obtains full agreement for $K\in\{3,5\}$ and full disagreement for $K\in\{10,20,30\}$. 

\subsection{Robustness with respect to the choice of $\alpha$}
\begin{figure*} \centering
  \includegraphics[width=0.7\textwidth]{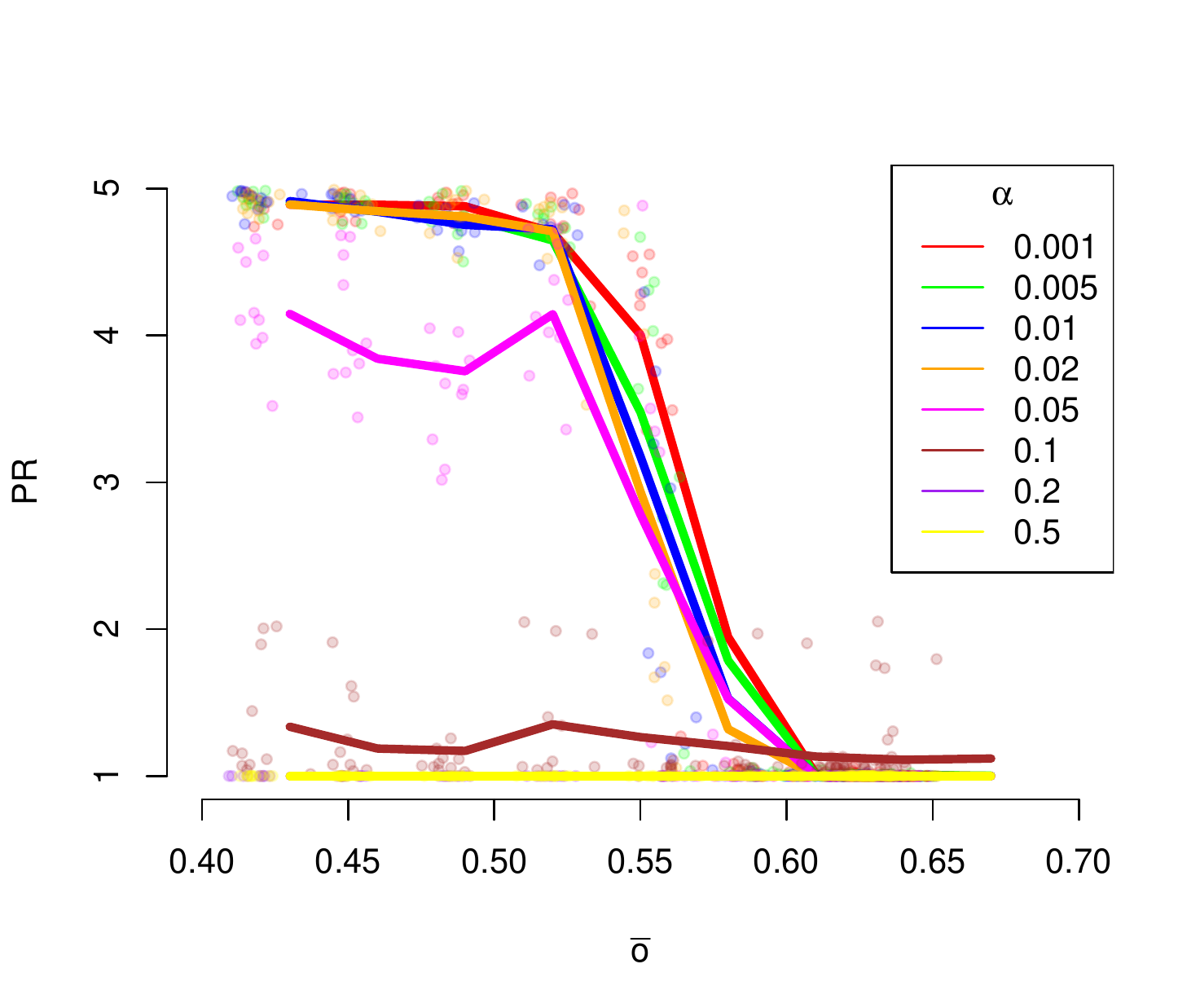}
  \caption{Effective number of clusters obtained for different initial
    conditions and $\alpha$ values for $K=5$.} \label{alpha}
\end{figure*}
The role of the value of the parameter $\alpha$ is analysed to
identify the effect of changing agent flexibility. In figure
\ref{alpha} we report the effective number of clusters PR resulting
from simulations with different values of the flexibility $\alpha$,
for $K=5$. The same dependence of the results on the initial condition
observed in Section \ref{initSect} is conserved as long as $\alpha$ is
small enough ($\alpha<0.05$ in simulations with $K=5$), whereas when
$\alpha$ is too large ($\alpha>0.1$), the population converges to one
cluster, regardless of the initial condition chosen. These results
show that the model is robust with respect to $\alpha$, as long as the
change in opinion is not forced to be very large, which is what is
expected. A very large $\alpha$ favours agreement in the population,
since individuals that are very different can become very close on a
single interaction, while for similar individuals which disagree, the
change is not as drastic, since the difference between their opinions
is very likely to be smaller than $\alpha$ (and hence use the second
rule in equation \ref{interEq}). It is important to note that decreasing $\alpha$ further below a certain value (thus increasing
simulation times) is not necessary since results are very similar.

\subsection{Temporal patterns}\label{sec:temporal}

\begin{figure*} \centering
  \includegraphics[width=\textwidth]{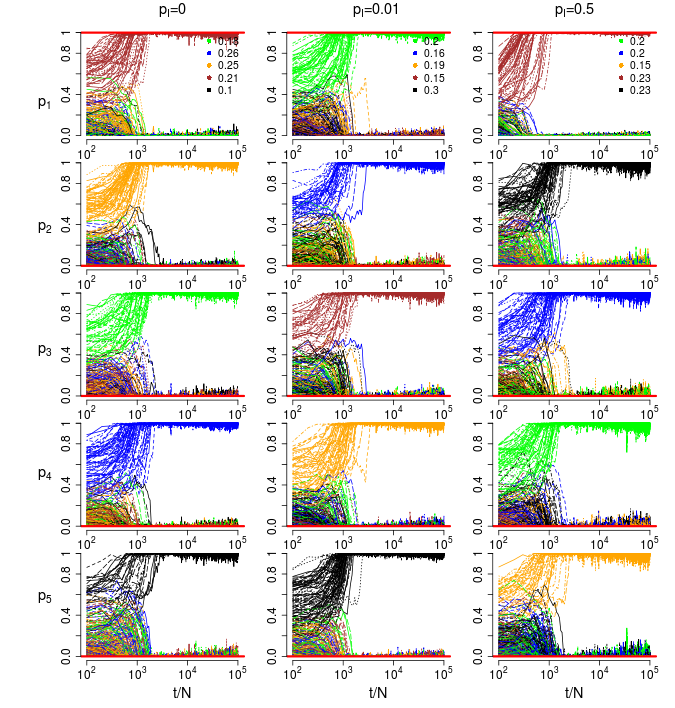} \caption{Opinion
    values for $K=5$, $N=300$, $\bar{o}=0.49$, $\vec I=[1,0,0,0,0]$. Each
    row corresponds to the positions in the opinion vector
    $\vec x=[p_1,p_2,p_3,p_4,p_5]$, while each column represents a
    different $p_I$. Individual opinions are coloured based on the
    cluster membership, with cluster sizes included in the legend. The
    information value is represented in red.  In this
    case five clusters are formed.} \label{dynI0S0} \end{figure*}
   
\begin{figure*} \centering
  \includegraphics[width=\textwidth]{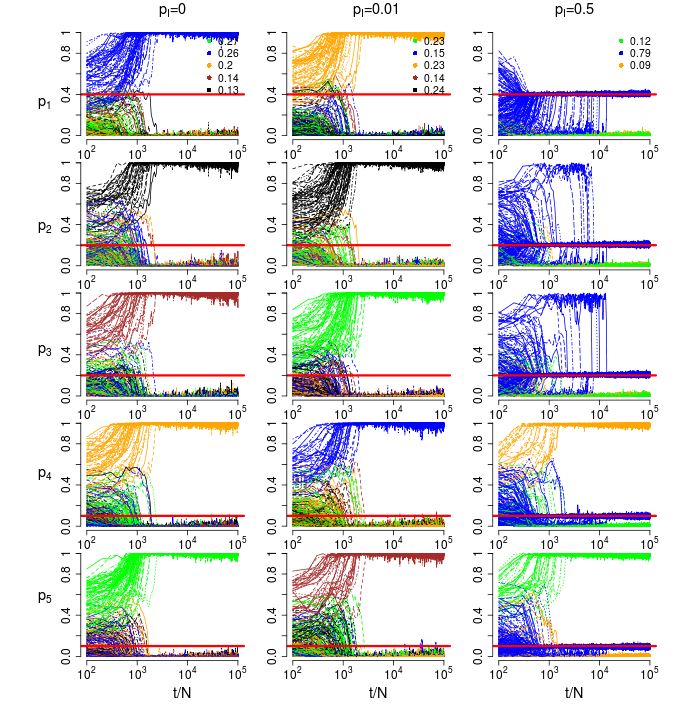} \caption{Opinion
    values for $K=5$, $N=300$, $\bar{o}=0.49$,
    $\vec I=[0.4,0.2,0.2,0.1,0.1]$. Each row corresponds to the positions
    in the opinion vector $\vec x=[p_1,p_2,p_3,p_4,p_5]$, while each column
    represents a different $p_I$. Individual opinions are coloured
    based on the cluster membership, with cluster sizes included in
    the legend. The information value is represented in red. In this case five
    clusters are formed for $p_I=0$ and $p_I=0.01$ and three for
    $p_I=0.5$.}
  \label{dynI1S0} \end{figure*}

\begin{figure*} \centering
  \includegraphics[width=\textwidth]{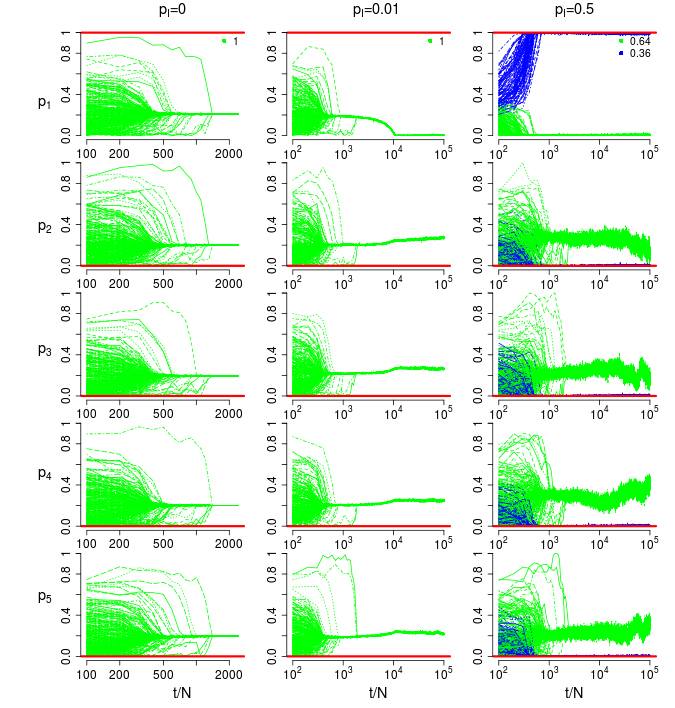} \caption{Opinion
    values for $K=5$, $N=300$, $\bar{o}=0.62$, $\vec I=[1,0,0,0,0]$. Each
    row corresponds to the positions in the opinion vector
    $\vec x=[p_1,p_2,p_3,p_4,p_5]$, while each column represents a
    different $p_I$. Individual opinions are coloured based on the
    cluster membership, with cluster sizes included in the legend. The
    information value is represented in red. In this case one cluster is formed for $p_I=0$ and $p_I=0.01$ and two for
    $p_I=0.5$.}
\label{dynI0S1}
\end{figure*}

\begin{figure*} \centering
  \includegraphics[width=\textwidth]{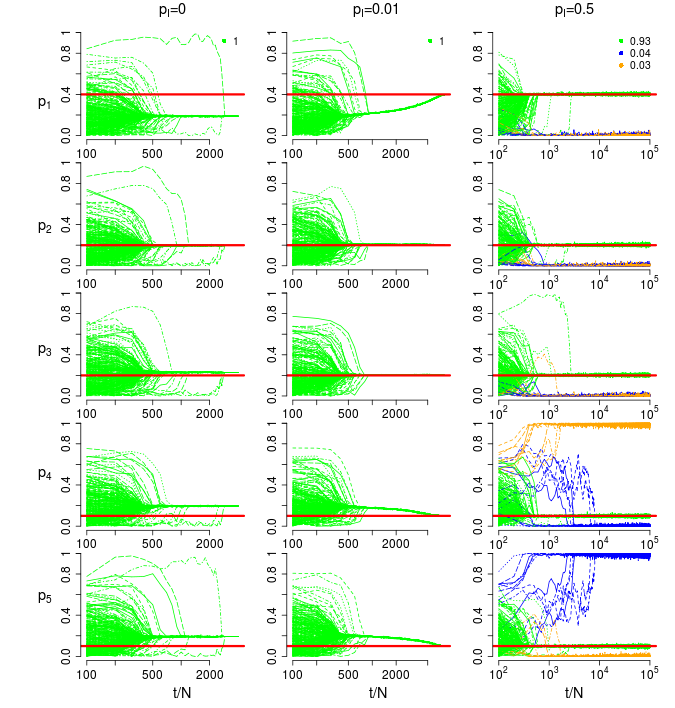} \caption{Opinion
    values for $K=5$, $N=300$, $\bar{o}=0.62$,
    $\vec I=[0.4,0.2,0.2,0.1,0.1]$. Each row corresponds to the positions
    in the opinion vector $\vec x=[p_1,p_2,p_3,p_4,p_5]$, while each column
    represents a different $p_I$. Individual opinions are coloured
    based on the cluster membership, with cluster sizes included in
    the legend. The information value is represented in red.  In this
    case one cluster is formed for $p_I=0$ and $p_I=0.01$ and three for
    $p_I=0.5$.}
\label{dynI1S1}
\end{figure*}
In order to get a further insight on the dynamics of the system, this
section discusses temporal patterns for two initial conditions
($\bar{o}=0.49$ and $\bar{o}=0.62$) and two information types
($\vec I=[1,0,0,0,0]$ and $\vec I=[0.4,0.2,0.2,0.1,0.1]$), each for $p_I
\in\{0,0.01,0.5\}$ and $K=5$. Figures
\ref{dynI0S0}, \ref{dynI1S0}, \ref{dynI0S1} and \ref{dynI1S1} show
single simulation instances for each $p_I$. Plots display the
evolution in time of each of the five elements of the opinion, for
every individual in a population of 300. Each row corresponds to one
position $p_i$ in the opinion vector, while each column represents a
different value of $p_I$. The value of the information is shown in
red. Individual opinions are displayed in colours corresponding to the
cluster they belong to at the end of the simulation (e.g.\ all green
lines correspond to individuals which cluster together). The relative
cluster sizes ($\frac{c_i}{N}$) are also shown at the top, as a
legend.

As figures show, opinions start in random position spanning the
interval $[0,1]$, and stabilise around a particular value. The system
never reaches a frozen state, with small fluctuations preserved even
after the clusters are formed (see Fig.~\ref{info5} for the effective
number of clusters in the different regimes). This is due to the
parameter $\epsilon$, which allows for agreement even when the overlap
between two individuals is zero, or disagreement even when the overlap
is one.

For the case of segregated initial populations, Figures \ref{dynI0S0}
and \ref{dynI1S0} show the formation of five clusters within the
population, for $p_I=0$. These five clusters are maintained for
extreme information, regardless of the value of $p_I$, due to the
segregation effect of such information. For milder external message,
on the other hand, the five clusters are only maintained when the
frequency of exposure is small. In this situation, however, although
the average overlap with the information is quite large (Figure
\ref{info5}), no individuals agree completely with the information.
Larger $p_I$ causes a large cluster to form around the external
information value, showing the cohesion effect of a mild message
(which appears only when the frequency of exposure is large enough).

Figures \ref{dynI0S1} and \ref{dynI1S1} show similar graphs for a
compact initial population. When no information is present, all
individuals form one cluster. Plots for $p_I=0.01$ validate the
explanation of the total agreement/disagreement observed in Figures
\ref{info5} and \ref{clusterSizes5}. Specifically, low exposure to the
information allows the population to initially form one cluster,
similar to no exposure, which is afterwards slowly affected by the
external message. When this is extreme (Figure \ref{dynI0S1}), the
cluster shifts away from the external information, since it is too
dissimilar. On the contrary, when the information is mild, the cluster
moves towards it resulting in complete agreement (Figure
\ref{dynI1S1}). For $p_I=0.5$, the group, determined previously by the
initial condition, does not form, and part of the population adheres
to the information directly. For extreme information, the cluster
overlapping with the information is small, while for the mild message,
this dominates the population.

All in all, we observe that a small $p_I$ allows for dynamics to be
determined by the initial condition at the beginning of the system's
evolution. Clusters are influenced by information after they are
formed: these move away or close to the information value, depending
on the cluster overlap with the external input. For larger $p_I$, the
initial overlap of each individual with the external information is
important: individuals who are far away form additional clusters that
are then too distant from the rest of the population and from the
information to be attracted back.

\subsection{Biased population}
\begin{figure*}
\centering
\includegraphics[width=\textwidth]{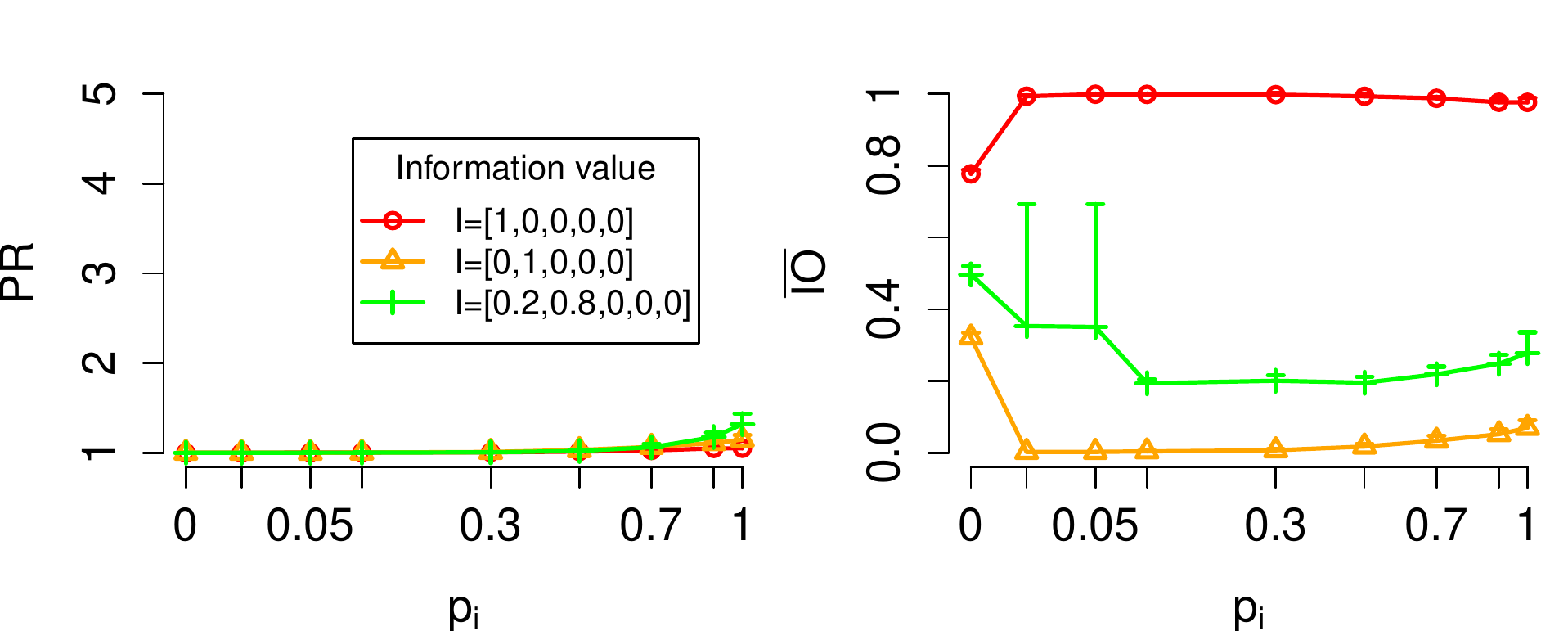}
\caption{Effect of extreme information on a biased population, with a positive bias for the first opinion choice (average opinion value for the population around $b=[0.4,0.15,0.15,0.15,0.15]$). The individual points correspond to $p_I\in\{0,0.01,0.05,0.1,0.3,0.5,0.7,0.9,1\}$.}
\label{biasedPop}
\end{figure*}

In the simulations presented until now, the population was random,
i.e., not biased, on average, towards any opinion. Information, on the
other hand, was considered extreme if it promoted strongly one of the
possible choices. Extreme information was shown to have limited effect
on such populations. However, there can be situations where the
initial population is already biased towards one of the possibilities,
in which case, extreme information can still have a large effect.
Figure \ref{biasedPop} shows the effect of different information types
for such a biased population. It is obvious that if the bias in the
population coincides with the opinion that the information promotes,
then even an extreme information can induce complete (for low $p_I$)
or wide (for larger $p_I$) agreement. However, if the choice promoted
by extreme information is different, then all individuals disagree.
Agreement with information, in the latter situation, increases when
information becomes milder. This shows that it is the similarity of
the information to the original population that determines its
success, and not the information structure in itself. In these
conditions, the previous observations for extreme information can be
extended to any situation, if we consider information ``extremism'' as a
feature dependent on the state of the population, i.e., information
with low overlap to the population.

\subsection{Population size}
   \begin{figure*}
\centering
\includegraphics[width=0.8\textwidth]{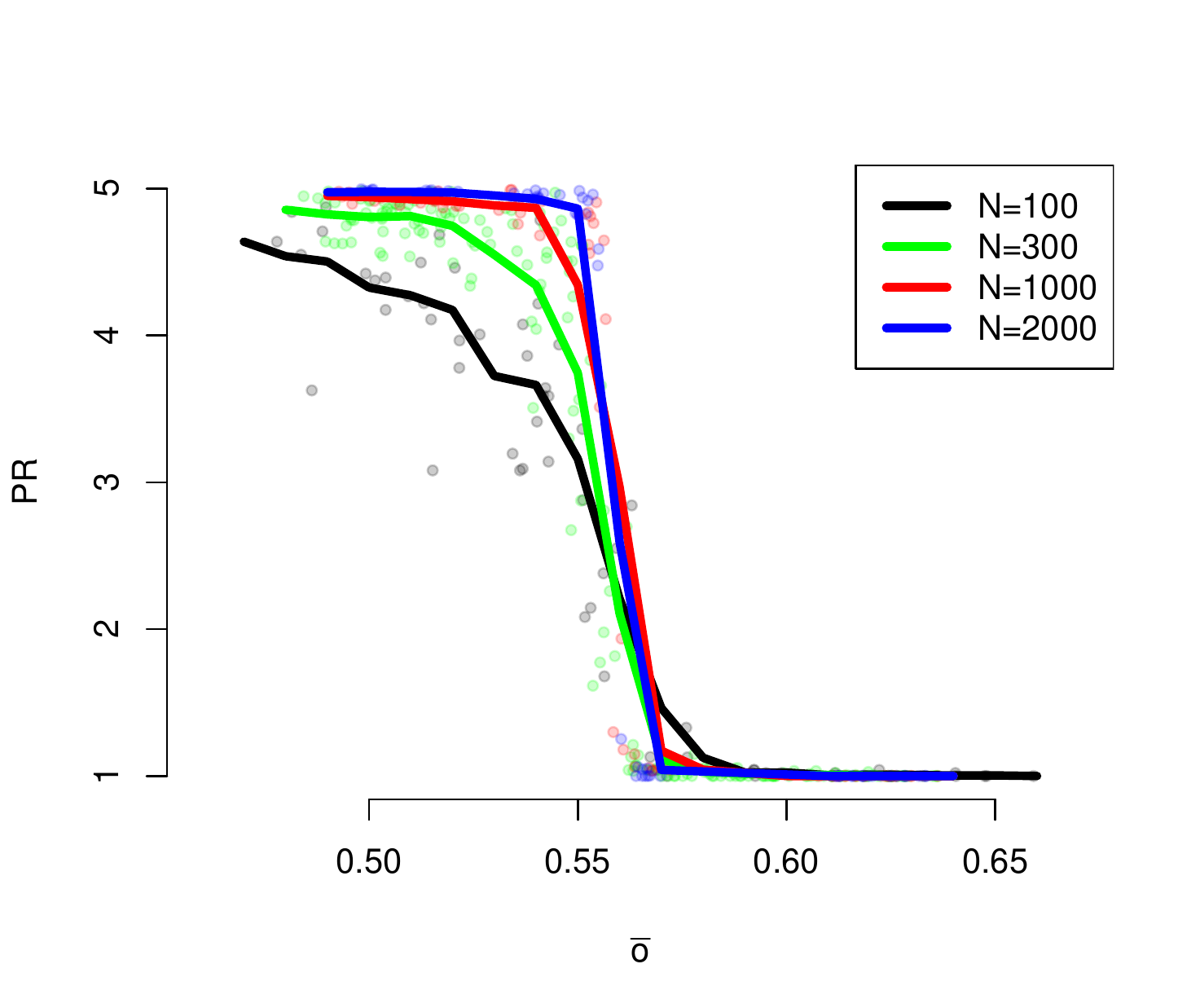}
\caption{Effect of the initial condition for different population sizes and $K=5$. Dots represent values for individual runs, while the lines show average number of clusters (computed by binning). }
\label{scalingInit}
\end{figure*}
All previous simulations presented have been performed using a limited
population size, i.e., $N=300$. Some effects seen in the results can be
artifacts of the small population size, hence in this section we are
analysing some of the previous simulation settings for increasing
population sizes i.e., $N \in \{100,300,1000,2000\}$. Due to increased
running times for larger populations, only the situation $K=5$ has
been studied, with ten instances for each parameter set.

A first such analysis involves the effect of the initial condition,
i.e., the average initial overlap in the population. Figure
\ref{scalingInit} shows the number of clusters for different instances
with various initial $\bar{o}$. This shows that the transition between
$K$ clusters obtained for low $\bar{o}$ and one cluster for large
$\bar{o}$, observed in Section \ref{initSect}, is conserved for
different populations sizes. Further this transition becomes steeper
as $N$ increases, with less frequent intermediate values at the
transition point. This indicates that the intermediate number of
clusters obtained for $N\in{100,300}$ could be an artifact of the
small population size; however, the curves overlap very well,
especially for larger $N$. These observations indicate a low
dependence between the effect of the initial condition and the
population size. As long as the population is not extremely small, (as
seen for $N=100$, where the curve seems more different than the
others), results do not change much, at least qualitatively, by
increasing the population size.

   \begin{figure*}
\centering
\includegraphics[width=\textwidth]{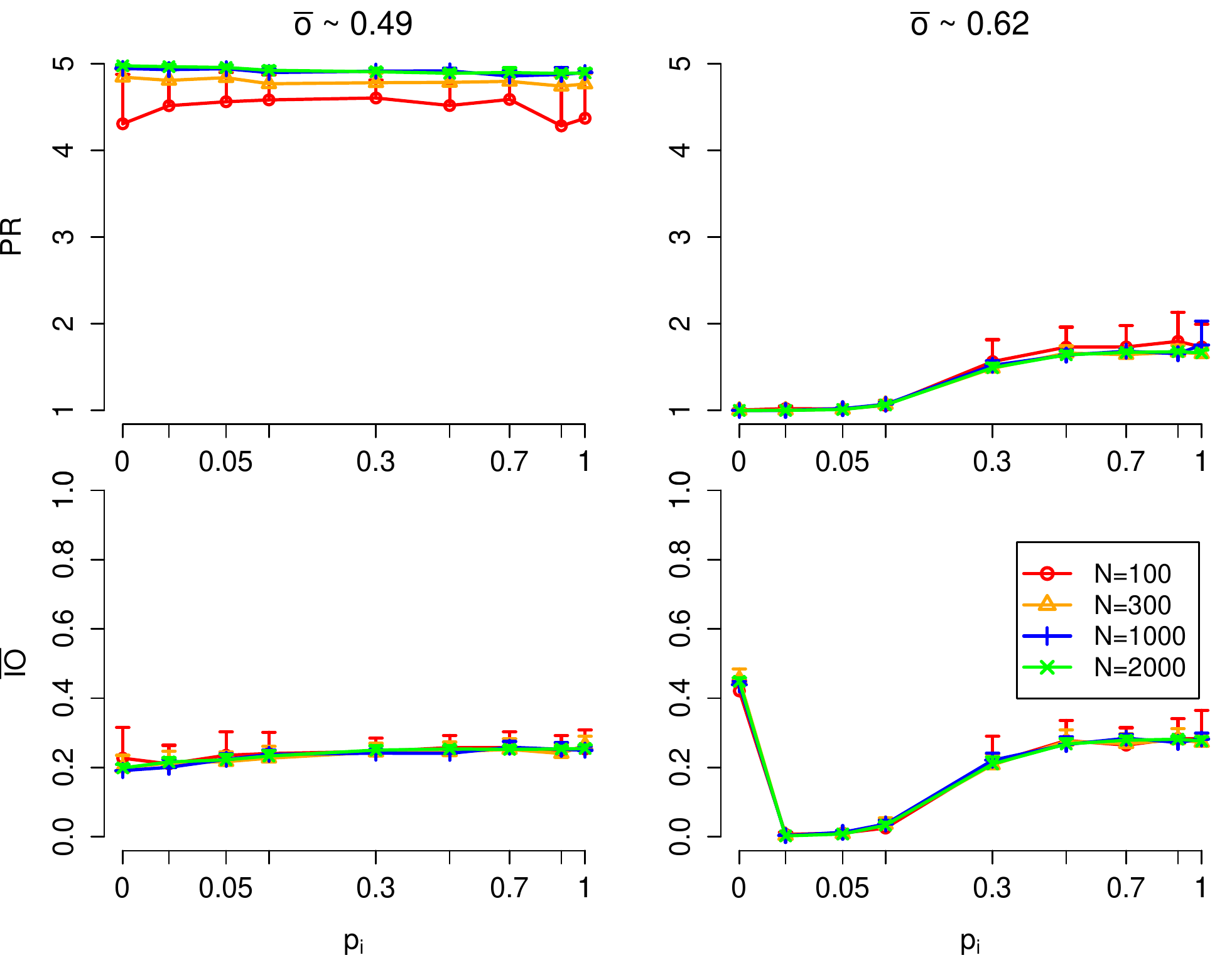}
\caption{Effect of extreme information ($I=[1,0,0,0,0]$) on K=5 and two different initial conditions, for different population sizes. The individual points correspond to $p_I\in\{0,0.01,0.05,0.1,0.3,0.5,0.7,0.9,1\}$.}
\label{scalingInfo}
\end{figure*}

In a similar fashion, the external information effect has been studied
for different population sizes, for $K=5$. Figure \ref{scalingInfo}
displays the average number of clusters and the final average
$\overline{IO}$ for four population sizes, $I=[1,0,0,0,0]$ and two
initial conditions (compact and dispersed initial population). This
shows an extremely good overlap between the curves corresponding to
each population size, indicating, for external information as well,
that results are not influenced by $N$. Cluster sizes, in the case of
segregated initial population ($\bar{o}\sim0.49$), become more uniform
as the population grows, i.e., PR value is closer to 5. This suggests
that, for an infinite such population, extreme external information
would produce equally sized clusters in the population.
\begin{figure*} \centering
  \includegraphics[width=\textwidth]{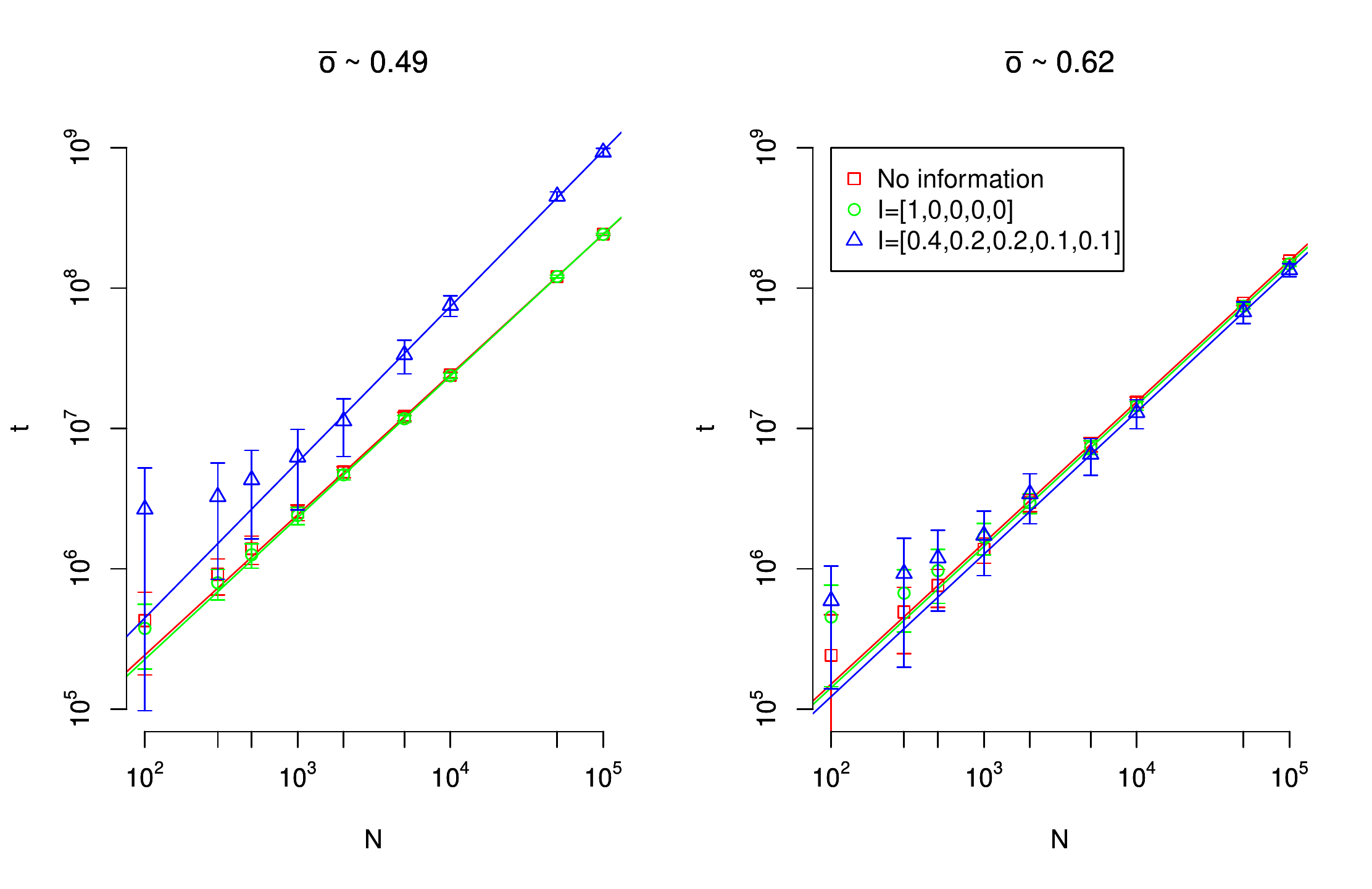} 
\caption{Relaxation
    times for two initial conditions (left and right plots,
    respectively) and two information types, vs. $N$.
    Times shown represent the number of pairwise interactions. Points
    represent averages over 100 simulation runs, with error bars
    corresponding to one standard deviation from the mean. The
    continuous lines show distributions of $t$ for large population
    sizes, fitted to the simulation times for $N\geq5000$. The
    distributions are of the form $t=\alpha N^\beta$, with $\beta=1$
    (i.e., linear dependency) for all situations except
    $\bar{o}\sim0.49$ and $\vec I=[0.4,0.2,0.2,0.1,0.1]$, when $\beta=1.11$
    (left panel, blue line).}
\label{times}
\end{figure*}

Further, we study relaxation times depending on $N$ and information
type. Since opinions do not converge to one value during simulations,
relaxation time is defined as the number of updates (total number of
interactions) required to obtain stable clusters and information
agreement, even though opinions will still fluctuate. Figure
\ref{times} displays the scaling, averaged over $100$ simulation runs,
of the relaxation times vs. $N$ for two information types (with
$p_I=0.5$) and two initial conditions. This shows that, in general,
clusters form faster when the initial population is more compact
(right vs. left plot). The scaling of the relaxation time with $N$
features a linear behaviour in all cases. An exception is a segregated
population exposed to mild information, where the scaling is slightly
more than linear with a fitted exponent $\sim 1.11$. Finite-size
effects are visible for small populations sizes, which features also
larger error bars. These finite-size effects can be explained by the
sparser sampling of the opinion space during initialisation, for small
populations.

The effect of external information on the relaxation time is also
displayed in Figure \ref{times}. The two initial conditions show
different trends. For a segregated starting point (left panel) the two
types of information have opposite effects on the relaxation time.
Specifically, exposure to mild information (which, as we saw before,
has a cohesive effect) increases significantly the relaxation time,
due to the contrasting effects of the initial condition and
information type (i.e., segregation vs cohesion). 
% On the other hand,
% times for extreme information are lower than those for the situation
% when no information is present, since both the initial condition and
% the information have segregating effects. 
For a compact initial condition (right panel) both information types
have the same effect.
% When $N$ is small, they increase relaxation times, while when the
% population becomes large enough, the time required is slightly
% shorter than when there is no exposure to information.

\section{Conclusions}

A new model for opinion dynamics was introduced, considering the
internal probability of individuals to choose between several discrete
options, i.e., with opinions represented as a vector of continuous
values with unity sum. Both attractive and repulsive dynamics are
considered, through a standalone mechanism, i.e., based on pairwise
similarity and not introducing further model parameters. An important
feature of the model is the ability to expose the system to modulated
external information, so that more possibilities can be promoted 
(e.g.\ by mass media).

Numerical results showed that extreme information causes fragmentation
and has limited success, while moderate information causes cohesion
and has a better success in attracting individuals. This coincides to
the success of marketing or election campaigns, where coalitions and
milder messages are more effective. Additionally, information success
is maximised when individuals do not interact too much with the
information, showing the importance of the social effect in
information spreading. An important factor driving the capacity of
information to influence the population appears to be the initial
similarity to the agents. This is a known fact in devising marketing
strategies, for instance, where information is displayed in a way
appealing to the target audience.

Similar effects of external information have been observed previously
for other models (either discrete or non-vectorial opinions). For
instance, \cite{Gargiulo2008a} observed that aggressive media
campaigns are not effective, using the Deffuant model with external
information, and that individuals should be exposed to the external
influence gradually in order to optimise its success. Further, in
\cite{Hegselmann2006,Kurz2011}, for the Hegselmann-Krause model, it
was shown that extreme information causes formation of antagonistic
clusters, while mild messages are more successful. A segregation
effect from external information has been also observed for the
Axelrod model \cite{Gonzalez-Avella2010,Peres2011,Peres2010}, similar
to our observations for extreme information (information in discrete
models, such as Axelrod, is equivalent to the extreme information in a
continuous model, i.e., promoting one option only).

The initial condition proved to be of large importance in the
dynamics, with compact populations resulting in one cluster, while
less compact starting points yielding more groupings. A similar large
effect of the initial condition has been recently shown in
\cite{Carro2012}, for the Deffuant model. Additionally, for our model,
the decrease in the transition point with $K$ indicated that agreement
is easier to obtain for a larger number of choices (with no external
information). This is similar to the findings for the model in
\cite{Lorenz2007}, which shares similarities to our approach. This
uses a bounded confidence threshold $d$, and it was shown that the
critical value of $d$, for which complete agreement is obtained,
decreases as $K$ increases. In our model, bounded confidence is not
present, however the initial condition plays a similar role in
determining the number of clusters, and we observe a very similar
effect of K on the critical $\bar{o}$ for which complete agreement is
obtained.

A scaling analysis showed that results for both the effect of the
initial condition and external information are robust with respect to
the population size. Additionally, relaxation time (total number of
updates), was shown to become linear in the population size as the
size increases.

Several further analyses of the model presented are envisioned for the
future. These include changing the dynamics to allow individuals to
interact on more than one opinion choice and studying different social
network topologies (here, all results are presented on a complete
network). Additionally, application to real data, in order to simulate
observed social processes and be able to make predictions, is required
to further validate the model. In the context of the EveryAware
project \cite{everyaware}, the model will be also applied to simulate
behavioural and opinion changes on environmental issues, based on
subjective data which will be collected during test cases.

\begin{acknowledgements}
This research has been supported by the EveryAware project
funded by the Future and Emerging Technologies program (IST-FET) of
the European Commission under the EU RD contract IST-265432 and the
EuroUnderstanding Collaborative Research Projects DRUST funded by the
European Science Foundation
\end{acknowledgements}

\bibliographystyle{spmpsci}      % mathematics and physical sciences
\bibliography{refs}   % name your BibTeX data base

\end{document}